\RequirePackage{etoolbox}
\newtoggle{daconf}

\iftoggle{daconf}{%
  \documentclass[12pt,conference,compsoc,onecolumn]{drivingassessment}
  \usepackage[round,sort,authoryear]{natbib}
  \bibliographystyle{abbrvnat}
  \setcitestyle{authoryear}
  \setlength{\bibsep}{6pt plus 0.3ex}
}{
  \documentclass[conference]{IEEEtran}
  \usepackage[square,sort,numbers]{natbib}
  \bibliographystyle{abbrvnat}
}

\RequirePackage{snapshot}

\makeatletter
\let\MYcaption\@makecaption
\makeatother
\usepackage[font=footnotesize]{subcaption}
\makeatletter
\let\@makecaption\MYcaption
\makeatother

\usepackage{subcaption}
\usepackage{cuted}
\usepackage{balance}

\usepackage{moreverb}
\usepackage{verbatim}

\usepackage{color}

\usepackage{parskip}

\usepackage{latexsym}
\usepackage{threeparttable}
\usepackage[pdftex]{graphicx}

\usepackage[sharp]{easylist}

\usepackage{multirow}

\usepackage{inconsolata}
\newcommand{\weblink}[1]{\path{#1}}

\usepackage[usenames,dvipsnames,table]{xcolor}

\usepackage{amsmath}
\usepackage{amsfonts}
\usepackage{amssymb}
\usepackage{amsthm}

\usepackage{algorithmic}
\usepackage{algorithm}
\usepackage{cases}

\renewcommand{\eqref}[1]{(\ref{eq:#1})}
\newcommand{\secref}[1]{\S\ref{sec:#1}}
\newcommand{\figref}[1]{Fig.~\ref{fig:#1}}
\newcommand{\tabref}[1]{Table~\ref{tab:#1}}

\usepackage{fancyhdr}
\fancypagestyle{firststyle}
{
  \fancyhf{}
  
   \fancyfoot[L]{
     \IEEEauthorrefmark{4} Corresponding author: \texttt{fridman@mit.edu}
   }
}

\begin{document}

\title{Dynamics of Pedestrian Crossing Decisions\newline Based on Vehicle Trajectories\newline in Large-Scale Simulated
  and Real-World Data}
  
\newcommand{\authorspace}{\hspace{0.2in}}
\author{
  \IEEEauthorblockN{
    Jack Terwilliger\IEEEauthorrefmark{1} \authorspace
    Michael Glazer\IEEEauthorrefmark{1} \authorspace
    Henri Schmidt\IEEEauthorrefmark{1} \authorspace
    Josh Domeyer\IEEEauthorrefmark{2}\\
    Heishiro Toyoda\IEEEauthorrefmark{2} \authorspace
    Bruce Mehler\IEEEauthorrefmark{1} \authorspace
    Bryan Reimer\IEEEauthorrefmark{1} \authorspace
    Lex Fridman\IEEEauthorrefmark{1}\IEEEauthorrefmark{4}
  }\\
  \IEEEauthorblockA{\IEEEauthorrefmark{1}Massachusetts Institute of Technology}
  \IEEEauthorblockA{\IEEEauthorrefmark{2}Toyota Collaborative Safety Research Center}
}


\maketitle

\begin{strip}
  \centering
  \captionsetup{type=figure}
  \begin{subfigure}[b]{0.48\textwidth}
    \includegraphics[width=\textwidth]{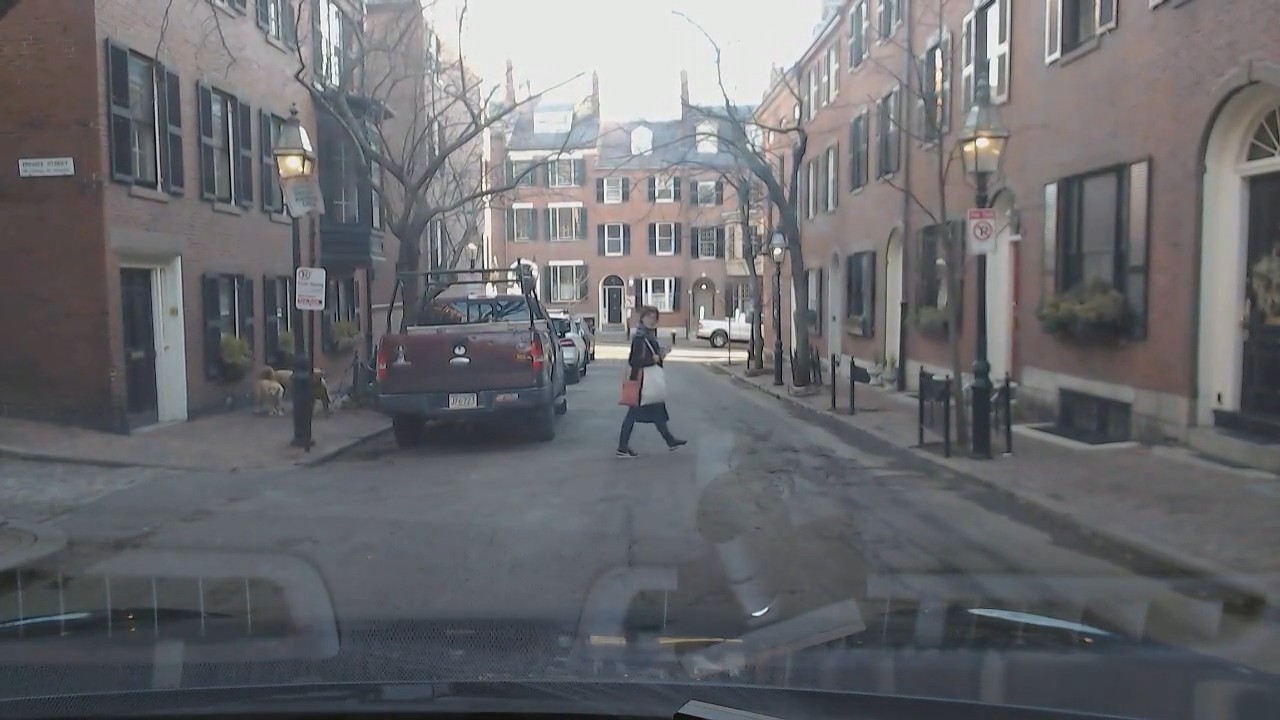}
    \caption{Example video frame of the forward roadway.}
    \label{fig:avt-frame}
  \end{subfigure}\hspace{0.15in}
  \begin{subfigure}[b]{0.48\textwidth}
    \includegraphics[width=\textwidth]{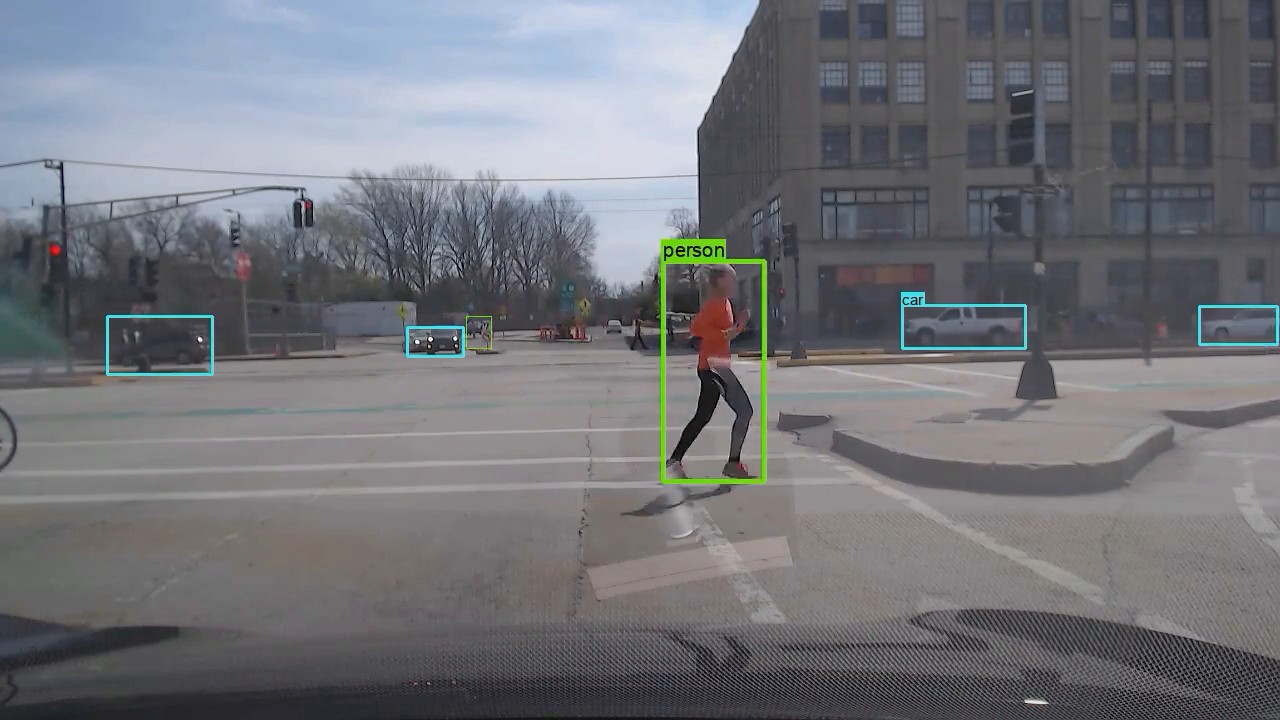}
    \caption{Pedestrian and vehicle detection via YOLO v3.}
    \label{fig:avt-detection}
  \end{subfigure}
  \caption{Example video frame and detection of vehicles and pedestrians from the MIT-AVT naturalistic driving dataset
    \cite{fridman2018avt}.}
  \label{fig:example}
\end{strip}

\begin{abstract}%
  Humans, as both pedestrians and drivers, generally skillfully navigate traffic intersections. Despite the uncertainty, danger, and the non-verbal nature of communication commonly found in these interactions, there are surprisingly few collisions considering the total number of interactions. As the role of automation technology in vehicles grows, it becomes increasingly critical to understand the relationship between pedestrian and driver behavior: how pedestrians perceive the actions of a vehicle/driver and how pedestrians make crossing decisions. The relationship between time-to-arrival (TTA) and pedestrian gap acceptance (i.e., whether a pedestrian chooses to cross under a given window of time to cross) has been extensively investigated. However, the dynamic nature of vehicle trajectories in the context of non-verbal communication has not been systematically explored. Our work provides evidence that trajectory dynamics, such as changes in TTA, can be powerful signals in the non-verbal communication between drivers and pedestrians. Moreover, we investigate these effects in both simulated and real-world datasets, both larger than have previously been considered in literature to the best of our knowledge.
\end{abstract}

\thispagestyle{firststyle}
\setlength{\footskip}{20pt}

\section{Introduction}\label{sec:introduction}

As experienced human drivers, we take for granted our ability to reason about pedestrians movements, intents, mental
models, and conflict resolution dynamics.  As pedestrian, vehicle passengers, and vehicle drivers, we quickly develop
the necessary perceptual capabilities such as foresight into whether a pedestrian is likely to cross the street and the
ability to communicate with pedestrians in explicit, non-verbal ways. As an illustration, consider a situation in which
someone is driving through a bustling street in downtown Boston. The driver spots a pedestrian on the sidewalk in the
middle of a city block walking towards the curb. She notices the pedestrian is looking in her direction. The pedestrian
pauses, but then the driver decelerates. The pedestrian then jaywalks (crosses outside a crosswalk) across the street in
front of the vehicle. While banal, this example encourages us to ask: (1) to what extent did the driver's influenced the
pedestrian's decision to cross and (2) how the driver was able to reason about the interaction. To design vehicle
automation that operates safely and efficiently in urban environments with an awareness of pedestrians, we will need
answers to the above questions. In this paper, we investigate (1) the relationship between vehicle trajectories and
pedestrian crossing decisions and (2) people's ability to update their estimates of a vehicle's time to arrival (TTA)
when vehicles accelerate.

Previous work has recorded the TTA between vehicles and pedestrians at the moment pedestrians begin to cross the
street. In 1953, \citet{moore1953pedestrian} first showed evidence that speed and distance influence when pedestrians
decide to cross and in 1955, \citet{cohen1955risk} began investigating TTA. More recently, \citet{brewer2006exploration}
found that 85\% gap acceptances (i.e., instances where pedestrians choose to cross) fall between 5.3 and 9.4
seconds. \citet{pawar2015pedestrian} provide convergent data, showing that, in developing countries, a similar
relationship exists between TTA and gap existence. While these studies have provided valuable information and models
about real-world crossing behavior, to design robust safety systems and vehicle automation, it's important to understand
how dynamics of trajectories, as opposed to a static notion of TTA, relate to pedestrian decision making.

Using simulators, previous works have measured and studied people's ability to estimate vehicle kinematics: the accuracy
of TTA estimation, the effects of velocity, and how we may use these estimates in deciding whether to cross the
street. \citet{petzoldt2014relationship} show that TTA estimations are influenced by vehicle speed and distance (e.g.,
pedestrians underestimate TTA at high velocities) and provide further evidence that pedestrians use TTA to decide
whether to cross. While these have been useful studies, the experiments have been limited to situations where vehicles
travel at constant velocities.

In order to understand pedestrian-vehicle interaction in greater depth we investigated behaviors both in a dynamic
real-world environment and through simulation that considers dynamic trajectories. We perform our analysis on two
large-scale datasets. The first is a real-world naturalistic driving dataset (see \secref{real-world}). The second is an
online simulated dataset (see \secref{online-simulation}). In \secref{results}, we present our results. In
\secref{conclusion}, we conclude with a discussion of applications to autonomous vehicle control algorithms and future
research directions.

\begin{figure}[ht!]%
  \centering
  \includegraphics[width=\columnwidth]{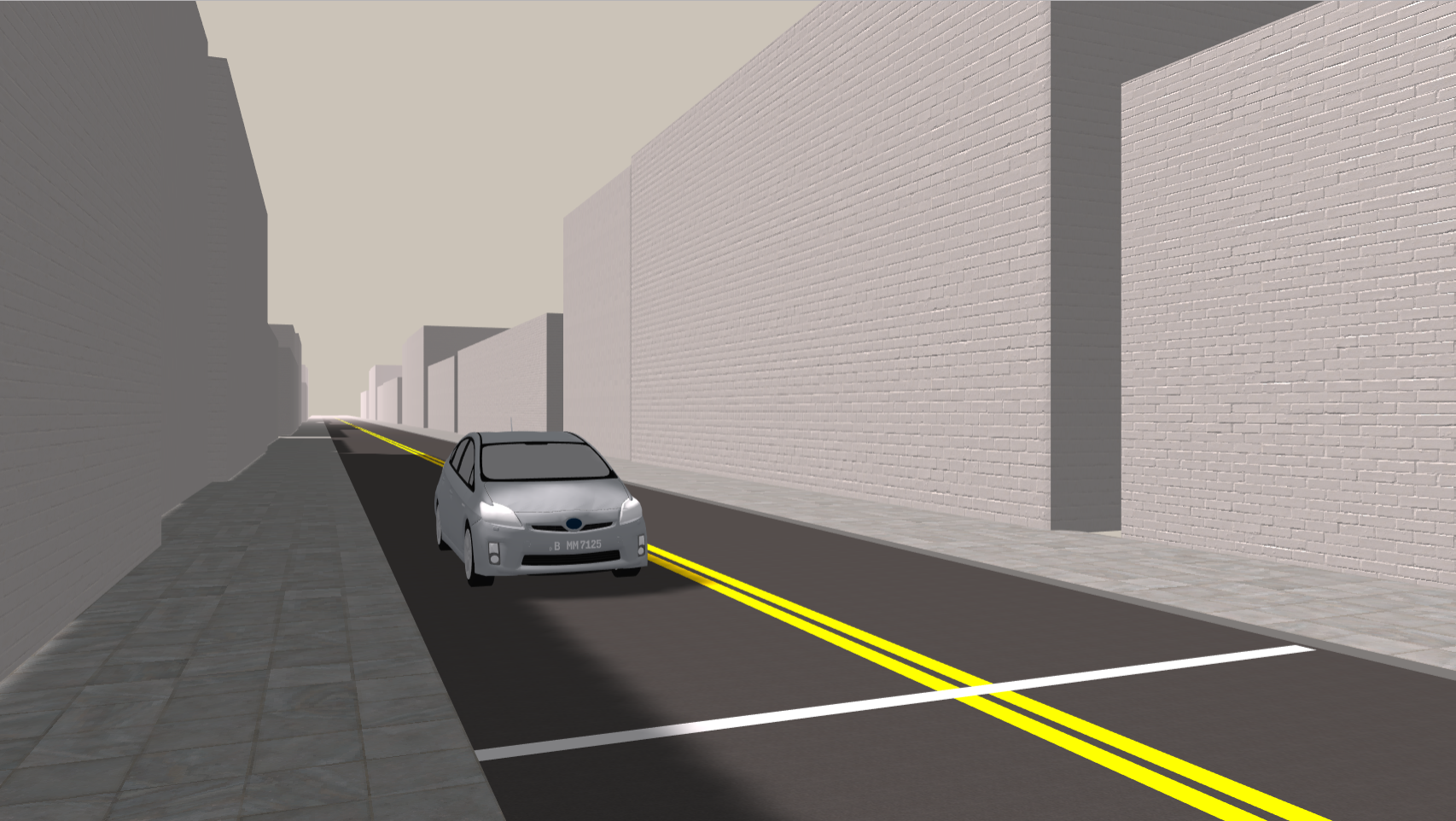}
  \qquad
  \caption{Screenshot from the online TTA estimation experiment.}%
  \label{fig:experiment}%
\end{figure}

\begin{figure*}[!ht]%
    \centering
    \includegraphics[width=\textwidth]{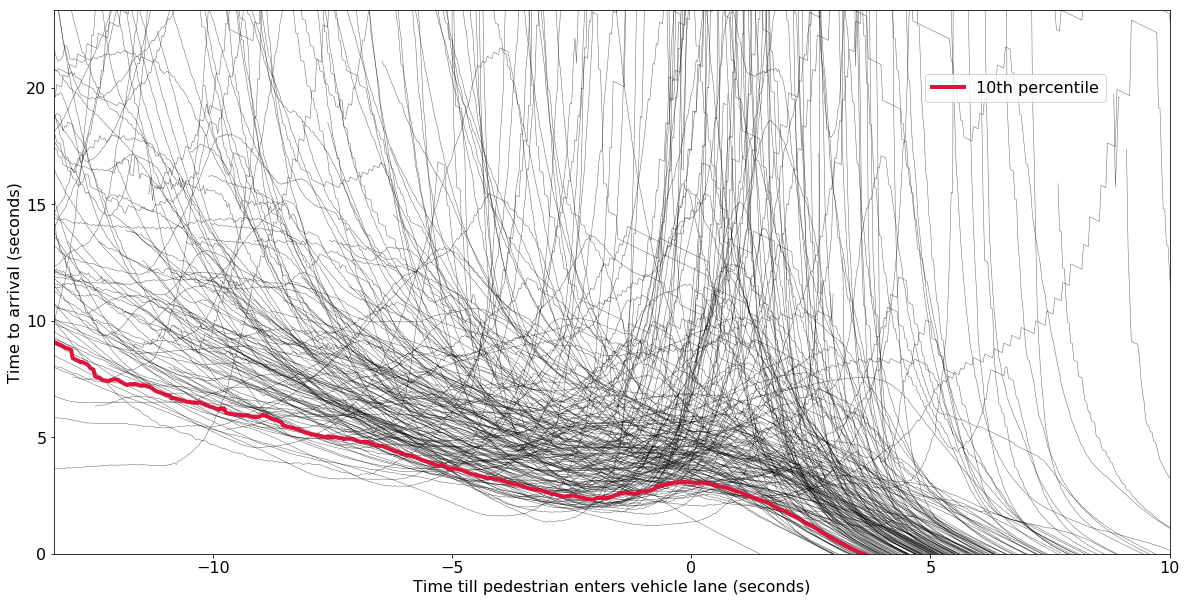}
    \caption{TTA $\frac{velocity}{distance}$ in seconds. All trajectories are aligned on the frame the pedestrian
      entered the path of the oncoming vehicle.}
    \label{fig:ttcreal-trajectories}%
\end{figure*}

\section{Methods}\label{sec:methods}

\subsection{Large-Scale Naturalistic Data Analysis}\label{sec:real-world}

Vehicle kinematics data originated from an approximately 200,000 mile subset of
the MIT-AVT naturalistic driving dataset
\citep{fridman2018avt,fridman2018humancentered}. The dataset includes data from
Greater Boston area drivers in vehicles equipped with automation technologies
throughout medium (1 month) and long-term (over a year) observations. This
dataset contains video, vehicle kinematics, and various messages from a
vehicle's systems. The video data include 720p 30fps video of (1) the forward
roadway, (2) the driver's face, and (3) the instrument cluster. Vehicle
kinematic data include odometer, speedometer, and steering angle information
recorded via the vehicle's CAN bus diagnostic port as well as GPS and IMU data
collected via an installed data collection system. Signals from a vehicle's
computer include previously mentioned kinematic data, whether the brake was
activated when a forward collision warning occurred etc.; a data collection
system recorded these signals via a CAN diagnostic port. For this study, we used
video of the forward roadway, vehicle kinematics, and GPS data. To ensure the
integrity of our analyses, all data were synchronized to video frames of the
forward roadway. See \figref{example} for examples of forward roadway video
frames and detections of pedestrians and vehicles.

In order to study how vehicle kinematics influence pedestrian behavior at
intersections, we needed to extract and annotate instances of short interactions
between drivers and pedestrians. Below, we outline a pipeline which involves (1)
a kinematics-based filter which excludes most highway driving (2) a computer
vision approach which extracts situations in which pedestrians likely crossed
the street, (3) a manual filter which selects only those interactions that fall
within a set of study criteria, and (4) a manual annotation tool for labeling
crossing-related events (e.g., entering the roadway, entering the path of the
approaching vehicle, etc.) and pedestrian body language (e.g., head orientation,
hand-waving, walking, standing, etc.). Note that the order of pipeline ensured
that more costly steps operate over the least amount of data.

\subsubsection{Kinematics-Based Filter}
  
To remove highway driving, the kinematics-based filter removes data in which
vehicles traveled faster than 50 mph. While this removed some non-highway
driving, we do not believe it significantly impacts the usefulness or
generalizability of our results, since pedestrian crossings most commonly occur
in urban settings where speeds are often much slower (approximately under 40 mph).

\subsubsection{Pedestrian Detection}

In order to extract sections of driving in which pedestrians likely crossed the
street, we, first, processed the remaining forward roadway video using YOLO v3
\citep{redmon2018yolov3} \citep{redmon2016you}, a real-time visual object
detection system. In the context of computer vision, object detection is the
problem of classifying and localizing (via bounding boxes) multiple objects in
an image. There are several practical advantages to YOLO v3, (a) YOLO v3 is a
deep learning based architecture which does not require manually crafted image
features, (b) YOLO v3 can process video 4x faster than comparable alternatives
(at 30fps on modern consumer hardware) \citep{redmon2018yolov3}, and (c) we were
able to detect the presence of more object classes than just pedestrian, which
provides value for future related research. We deployed a Darknet
\citep{darknet13} implementation on a computer cluster, in order to process
video in parallel.

After performing pedestrian detection in every frame, we used a heuristic for
selecting frames in which a detected pedestrian was likely to be crossing the
street. If a bounding box was found in the middle-third (horizontally) of the
frame, we flagged the frame as likely containing a crossing pedestrian. This
middle-third section served as a conservative approximation of the road region
in the scene. The overall heuristic approach performed very well at correctly
identifying crossing pedestrians and at filtering out non-crossing pedestrians.
The approach was validated by manually annotating a small subset of the detected
frames and measuring the false accept rate (FAR) and false reject rate (FRR) of
the heuristic selection approach. The middle-third of the video frame was
approximately the size needed to achieve a minimum equal error rate (EER).

We then extracted 30-second video clips of the detected pedestrian crossings: 20
seconds prior to the frame with a crossing pedestrian and 10 seconds after it.
If two videos overlapped, we combined them into one video.

\begin{figure*}[!ht]%
    \centering
    \includegraphics[width=4.5in]{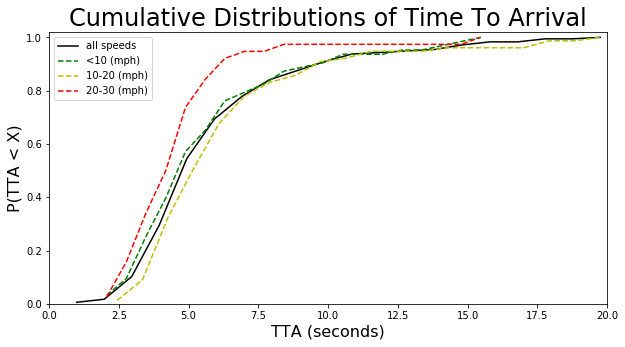}
    \caption{TTA $\frac{velocity}{distance}$ in seconds at the moment a pedestrian entered the path of the oncoming
      vehicle.}%
    \label{fig:ttcreal-cdf}%
  \end{figure*}
  
\subsubsection{Manual Filter}

In order to remove irrelevant data, we manually filtered the video clips
resulting from step (2). We define relevant data as video matching the following
criteria: (a) a pedestrian crossed as part of a group of less than 5; (b) the
lead pedestrian was visible when they entered the path of the vehicle; (c) the
instrumented vehicle was the lead vehicle; (d) the vehicle was moving before the
pedestrian crossed the street. We manually watched the videos and either
accepted or rejected them for consideration in the annotation process. To do
this, we built a simple OpenCV/Python tool to play video at 10x speed and
keep/remove interactions with key presses.

\subsubsection{Manual Annotation of Crossing Event Characteristics}

In order to label crossing-related events and pedestrian body pose, we
manually annotated the videos using a custom OpenCV/Python tool. All annotations
were of or relative to the lead pedestrian. Body pose included (a) whether a
pedestrian’s head was oriented toward or away from the driver or whether it was
oriented down, (b) whether the pedestrian was standing, walking, or running, (c)
whether the pedestrian waved at the vehicle. Crossing events included (a) when
the pedestrian entered the roadway, i.e. when the pedestrian stepped onto the
roadway (b) when the pedestrian entered the paths of the ego-vehicle, which may
occur after the pedestrian steps onto the road (c) when the pedestrians exited
the path of the ego vehicle, (d) when the pedestrian exited the roadway and (e)
when the vehicle crossed the path the pedestrian took to cross, i.e. the point
where the pedestrian’s and the vehicle’s paths crossed. Features of the
intersection included (a) whether the intersection occurred at a stop light, (b)
whether the intersection included a zebra crossing, (c) whether the pedestrian
was jay-walking.

\subsection{Simulator Experiment}\label{sec:online-simulation}

Our simulator experiment tested people's ability to estimate TTA under conditions when (1) vehicles approached at a
constant velocity and (2) vehicles approached while decelerating. A screenshot of the virtual environment is shown in
\figref{experiment}. This experiment was designed and conducted to supplement the large-scale real-world dataset of
pedestrian crossing in \secref{real-world} in order to analyze the nuance of vehicle trajectory dynamics as they relate
to pedestrian crossing decision. In real-world data, we cannot control either the pedestrians nor the vehicles, but
simple observe and analyze the kinematics of both. In the virtual environment, we can control the vehicle trajectory
and observe its effect on the pedestrian crossing decision.

\textbf{Design:}

In this experiment, we tasked participants with estimating the time to arrival (TTA) of a vehicle: or specifically, with
estimating when a vehicle would reach a white line painted across a virtual road. In each trial, after traveling some
distance, the vehicle disappeared before reaching the white line. This forced participants to estimate the TTA based on
prior kinematic information. We define the ground truth TTA as the time between the moment a vehicle disappeared and the
time it would arrive at the white line. We measured participants' estimated TTA by asking them to press the spacebar on
their computer when they thought the vehicle would reach the white line -- estimated TTA is thus the time between the
moment the vehicle disappeared and the time a participant pressed the spacebar.

In the first condition, a vehicle approached the white line at a constant velocity 3 seconds prior to disappearing. The
independent variables were the velocity of the vehicle and the ground truth TTA, i.e., the vehicles approached at
different speeds (5 mph \& 30 mph) and disappeared at different TTAs. The dependent variable was participants' estimated
TTA.

In the second condition, a vehicle approaching the white line at 30mph and decelerated to 0 mph, stopping at the white
line. At a predetermined distance after decelerating, the vehicle disappeared. The independent variables were the
deceleration ($-1.9m/s^2$ - $6.5m/s^2$) and the ground truth TTA. The dependent variable was participants' estimated
TTA. Because the vehicle always starts at 30 mph and ends at 0 mph, varying deceleration also varies the distance at
which the vehicle begins decelerating, according to kinematic laws. In this second condition, there is less freedom to
vary the ground truth TTA. This is because the vehicle must not disappear before it begins decelerating (if it were to,
participants would not perceive information necessary for estimating TTA).

In each condition, the vehicle reappears when participants press the space bar.  This provides feedback akin to a real
world situation in which a person estimates the time to arrival of a vehicle and later observes the actual time to
arrival as the vehicle reaches them. Within each condition, we show participants trajectories in random order.

\textbf{Implementation:} We ran this experiment on Mechanical Turk. Using three.js, a library utilizing WebGL, our tool
rendered, in real time, the virtual scene. While realtime rendering was not necessary for this experiment, as we could
have used pre-rendered videos, it may enable interactive experiments in the future, e.g., the vehicle reacts to
participant input.

\textbf{Participants:} A total of 66 people participated in the TTA experiment with 42 males and 24 females. To mitigate
the effects of poor render speeds, if during a trial, the frame-rate dropped below 30 fps, we removed the trial from
consideration. Additionally, to mitigate the effect of different screen sizes, when a participant's screen was narrower
than 1000px, the experiment prompted users to resize their window to continue.

\begin{figure*}[!ht]%
  \centering
  \begin{subfigure}[b]{0.48\textwidth}
    \includegraphics[width=\textwidth]{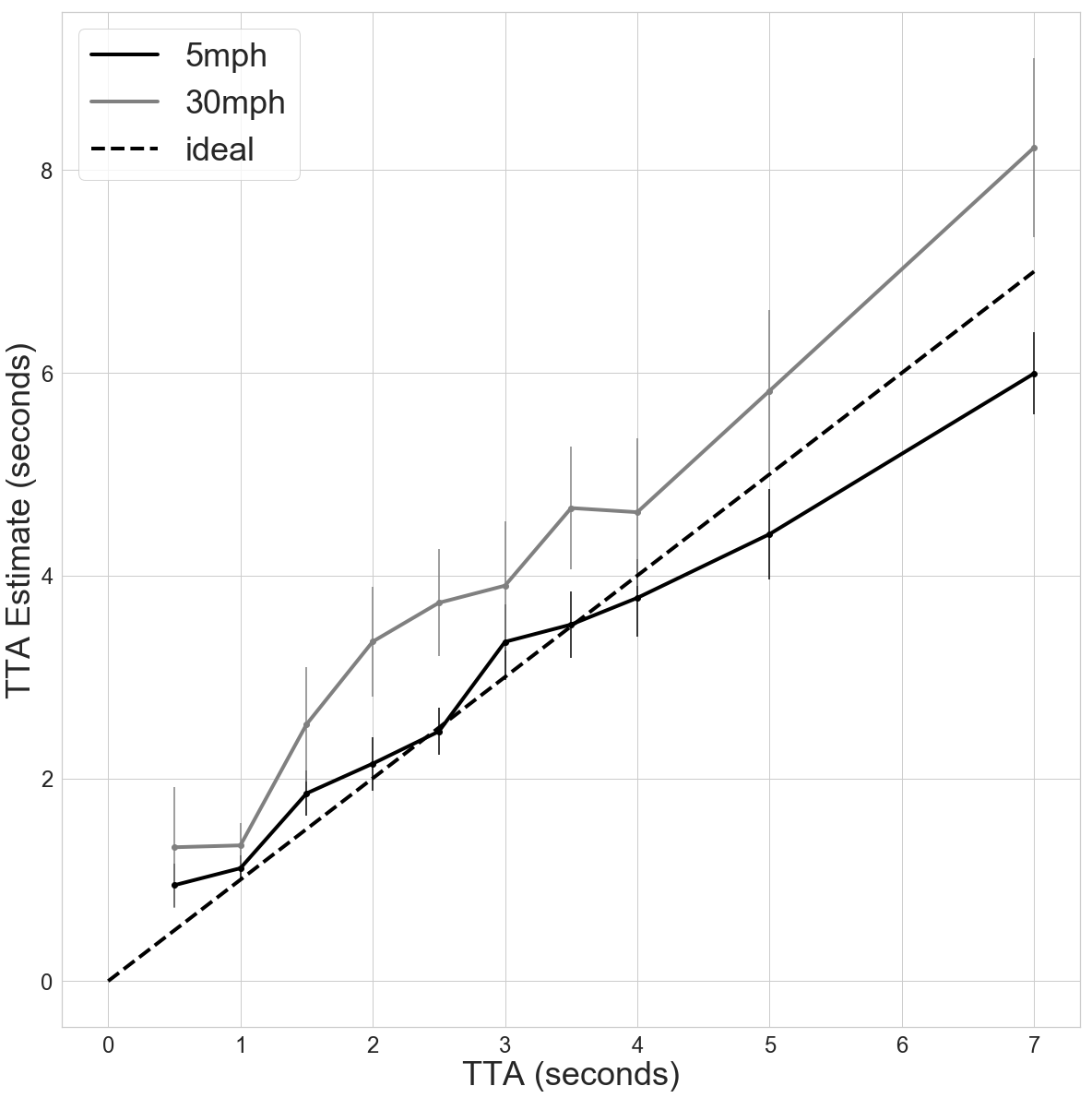}
  \end{subfigure}
  \hspace{0.1in}
  \begin{subfigure}[b]{0.48\textwidth}
    \includegraphics[width=\textwidth]{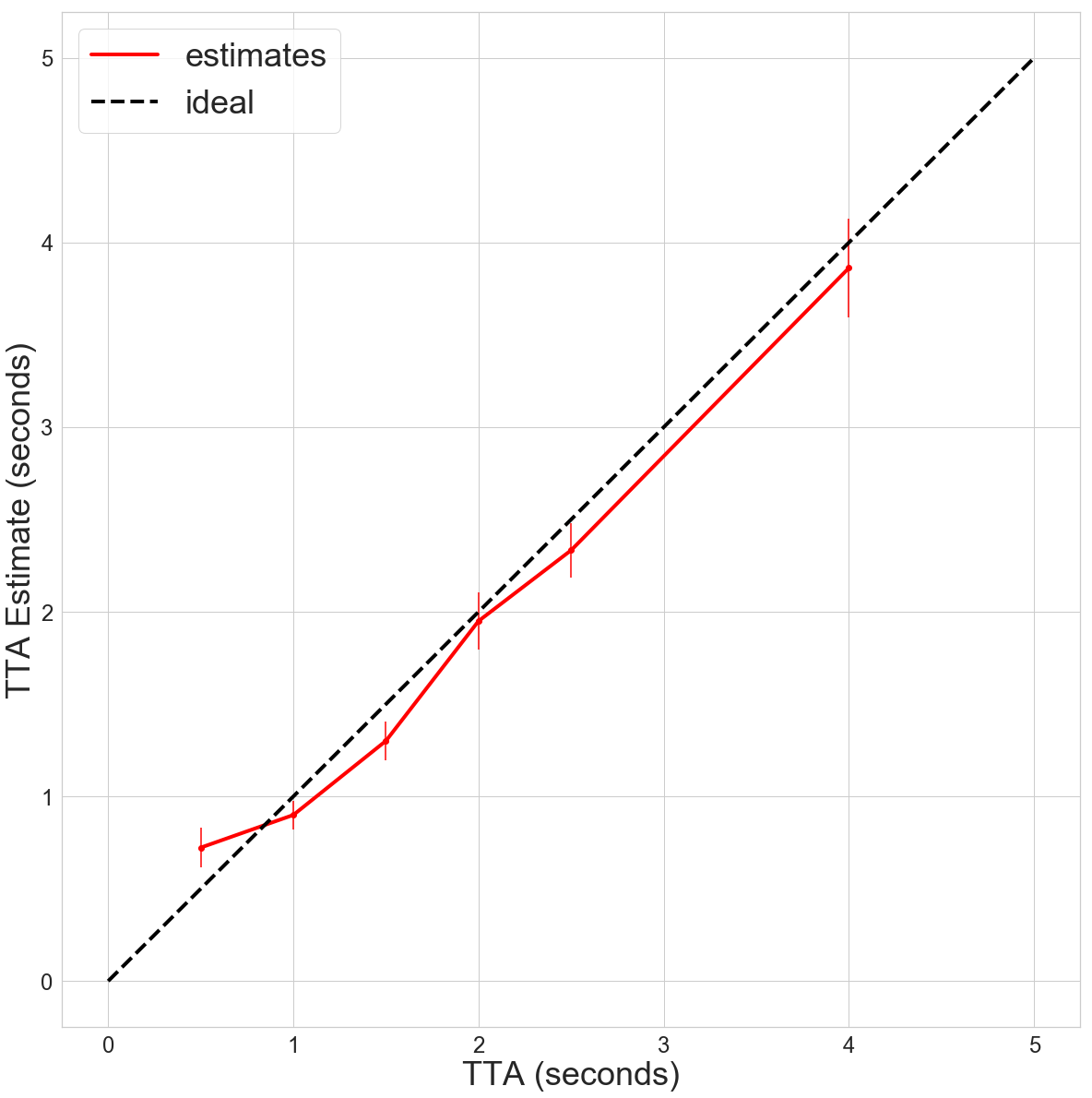}
  \end{subfigure}
  \caption{(Left) Participants' estimates of TTA of a vehicle traveling at constant velocities: 5mph (solid black) and
    at 30mph (solid grey). Participants overestimated TTA when the vehicle traveled at 30mph. Participants
    underestimated TTA when the vehicle traveled at 5mph when the car was far away.  (Right) Participants' estimates of
    TTA of a vehicle traveling at a constant velocity and then decelerating.}%
  \label{fig:ttcbyvel}%
\end{figure*}

\section{Results}\label{sec:results}

\subsection{Large Scale Naturalistic Data Analysis}

We now illustrate the characteristics of vehicle trajectories found ``in the
wild'' in situations where pedestrians chose to cross. Specifically, we show (1)
evidence that temporal dynamics influence pedestrian decision-making, and (2)
results convergent with \citep{petzoldt2014relationship} which suggest that,
while pedestrians use TTA when deciding whether or not to cross, they
underestimate the TTA at higher velocities.


(1) In \figref{ttcreal-trajectories} we show 284 vehicle trajectories (TTA over time) relative to the moment a lead
pedestrian entered the path of the vehicle.  While it may appear redundant to plot TTA over time, because vehicles
accelerate/decelerate as they approach, in order to accurately estimate the time they have to cross, a pedestrian must
update their estimates over time. We see a trend, 34\% of drivers slow the vehicle such that the time to collision
increases before the pedestrian steps in front of their vehicle. Here, TTA refers to a simple linear extrapolation of
vehicle kinematics, i.e.  $\frac{velocity}{distance}$. To normalize the data, we align each trajectory on the frame in
which an annotator determined a pedestrian entered the path of the oncoming vehicle. Though we are unable, with these
data, to ask the counterfactual ``what if the driver had not slowed down?'', these data suggest that, in real-world
situations, pedestrians tend only to cross when vehicles slow down such that the time the pedestrian has to cross
increases.

(2) In \figref{ttcreal-cdf} we show the empirical cumulative distributions of
TTA at the moment crossing pedestrians entered the path of the oncoming vehicle
N=195 (we removed cases where TTA was greater than 20). Performing a
Kolmogorov-Smirnov test between each category of vehicle speed indicates a
significant difference between when pedestrians cross the street in cases where
vehicles traveled between 10-20 mph and cases where vehicles traveled between
20-30 mph (D-statistic=0.15, p<0.05). The test does not indicate significant
differences between any other pair of vehicle speed categories see \tabref{kstest}.

\begin{table}[H]
  \centering
  \caption{Kolmogorov-Smirnov Test Table}
\begin{tabular}{lll}
\hline
Samples                        & D-Statistic & p-value \\ \hline
\textless{}10 mph \& 10-20 mph & 0.15        & 0.375   \\
\textless{}10 mph \& 20-30 mph & 0.22        & 0.145   \\
10-20 mph \& 20-30 mph         & 0.30        & 0.011*   \\ \hline
\end{tabular}\label{tab:kstest}
\begin{tablenotes}
  \item Results of a Kolmogorov-Smirnov Test between each pair of the three
    vehicle speed categories.
\end{tablenotes}
\end{table}

These results, taken from unconstrained real-world situations, provide strong
supplementary evidence, that pedestrians base their decision of when to cross on
TTA. We find, surprisingly, at higher speeds, pedestrians enter the lane with
less time than at lower speeds. According to \citep{petzoldt2014relationship},
pedestrians overestimate the TTA at higher speeds -- a result consistent with
other literature \citep{hancock1997time} \citep{sidaway1996time}. We note that
\citep{petzoldt2014relationship} did not find evidence that overestimating TTA
influenced gap acceptance. The Petzoldt \citep{petzoldt2014relationship} study
was conducted in a lab setting and the differences between our findings and
theirs may be the result of their participants becoming aware of and correcting
for their tendency to overestimate the TTA in a predictable environment.

\subsection{Simulator Experiment}

We now illustrate the results of how our participants were able to estimate TTA when a vehicle was traveling at a
constant velocity and when a vehicle was decelerating.

In \figref{ttcbyvel} (left), we show evidence that people overestimate TTA of vehicles traveling at higher
velocities. The plot shows the ground truth TTA (x axis) vs. participants' estimates of TTA (y axis). The dashed black
line (x=y) shows what an ideal estimator would look like. Estimates above the dashed line are over estimates; estimates
below the dashed line are under estimates. This data agrees with \citep{petzoldt2014relationship} that vehicle speed
influences TTA estimates. This suggests the source of our findings from naturalistic study (that pedestrians enter the
lane sooner under less TTA when vehicles are traveling at high speeds) is based on the perceptual bias -- to
overestimate TTA when vehicles are traveling at high speeds.

In \figref{ttcbyvel} (right), we show that people are sensitive to changes of speed and are able to rapidly update their
estimates of the kinematics of oncoming vehicles. As in the previous plot, this plot shows the ground truth TTA (x axis)
vs. participants' estimates of TTA (y axis). This demonstrates that, as expected, people are able to rapidly update
their estimates of the kinematics of oncoming vehicles. This result provides grounds for interpreting our findings that
drivers alter their trajectories as they approach pedestrians as a non-verbal signal, which pedestrians may use to infer
the intent of drivers.

\balance

\section{Conclusion}\label{sec:conclusion}

As more of the driving task becomes automated, we must deepen our knowledge of how pedestrians react to trajectories of
human-driven vehicles. Closing this knowledge gap is important for developing both effective autonomous motion planning
algorithms and communication protocols in a mixed fleet that includes vehicles controlled both by humans and machines.

Here we have shown evidence that (1) in real-world situations pedestrian decision-making is
biased -- they tend to give themselves less time when vehicles travel at faster speeds, (2) dynamics of vehicle
trajectories, namely increases in TTA, appear to serve as signals that it is safe to cross, and (3) that people can
update their estimates of TTA as vehicles change speed. While these results provide a pragmatic conclusion, that
automated technology ought to account for human bias to overestimate TTA at higher speeds, they also motivate the need
to further study of dynamic trajectories in order to understand pedestrian-driver interactions at a more nuanced level.

One limit to the real-world dataset considered in our work is the absence of situations in which pedestrians did not
cross. Future research will study factors of driver behavior that discourage pedestrians from crossing.  Additionally,
data from Boston may not necessarily generalize to other places.

We conclude that the everyday act of crossing the street is a nuanced dialogue between pedestrians and drivers. An
understanding of this dialogue requires an understanding of people's theory of mind -- at least in the specific context
of crossing the street. Other future work will explore pedestrian body language when attempting to cross the street and
explore whether pedestrians can infer driver intentions purely from kinematic information.

\section*{Acknowledgment}

This work was in part supported by the Toyota Collaborative Safety Research Center. The views and conclusions being
expressed are those of the authors and do no necessarily reflect those of Toyota.

\bibliography{pedestrian}

\end{document}